\documentclass[12pt,a4paper]{article}
\usepackage{epsfig}
\usepackage{color}
\usepackage[english]{babel}
\usepackage[latin1]{inputenc}
\usepackage{times}
\usepackage[T1]{fontenc}
\newcommand{\eps}{\varepsilon}		
\begin{document}
\vskip2cm
\begin{center}
{\Large{\bf{Maximal Net Baryon Density in the Energy Region Covered by NICA.}}}
\vskip1cm
{\bf J.\ Cleymans}
\\[1cm]
 UCT-CERN Research Centre and Physics Department, University of Cape Town, South Africa,\\
~\\
\bigskip

Abstract:\\
\end{center}
There  are several theoretical  indications that the energy region 
covered by the proposed  NICA accelerator in Dubna is an extremely 
interesting one.
We present a review of data obtained in relativistic heavy ion collisions and show that there
is a gap around 10 GeV where more and better precise measurements are needed.
The theoretical interpretation can only be clarified by covering this 
energy region.
In particular the strangeness content needs to be determined, data covering
the full phase space ($4 \pi$) would be very helpful to establish the thermal parameters of 
a possible phase transition. 

\newpage
~~~~~~~~~~
\vskip1cm
\section{Introduction}
Particle collisions at high energies produce large numbers of secondaries and it is natural to try
a statistical-thermal model to analyse these.  This type of analysis has a long 
and proud history~\cite{koppe,fermi,hagedorn} and led to the successful explanation of
$m_T$ scaling which is a natural consequence of such  models. The behaviour in the longitudinal direction 
was however very different and led many people to
discard the thermal model. \\
In  relativistic heavy ion collisions a new dimension was given to the model
by the highly successful analysis of particle yields, leading to the notion of chemical 
equilibrium which is now a  well-established one in the analysis of relativistic heavy ion collisions~\cite{cleymans-satz}. 
The early situation in 1999 is summarized in Fig.~1, with three points, showing a clear increase of the 
chemical freeze-out temperature, $T$,
with increasing beam energy and an accompanying decrease of the baryon chemical  potential $\mu_B$~\cite{becattini}.
\begin{figure}[htb]
\centerline{\epsfig{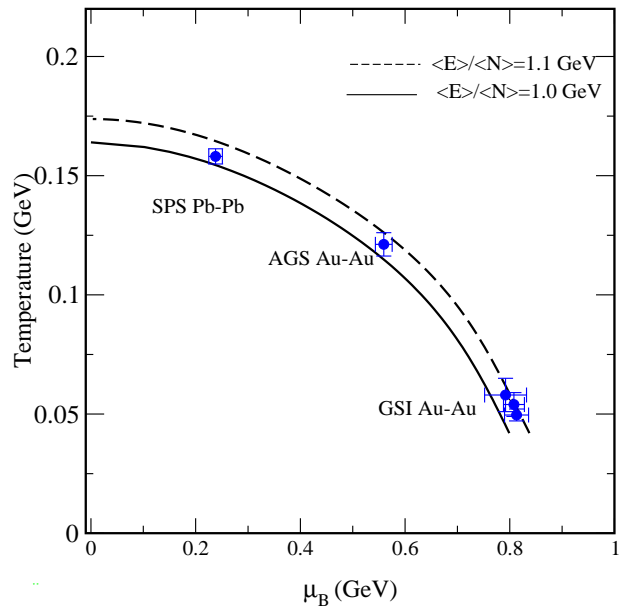}}
\caption{Chemical freeze-out temperature $T$ vs. the baryon chemical potential  at different beam 
energies together with curves corresponding to a fixed ratio of energy per hadron divided by
total number of hadrons in the resonance gas before decay of resonances.}
\label{e_1999}
\end{figure}
The situation improved  substantially in the following decade~\cite{pbm,manninen,picha,takahashi} and now covers almost 
the complete curve as shown in Fig.~2. Note that a last substantial gap
still exists in the energy region to be covered by NICA.
\begin{figure}[htb]
\centerline{\epsfig{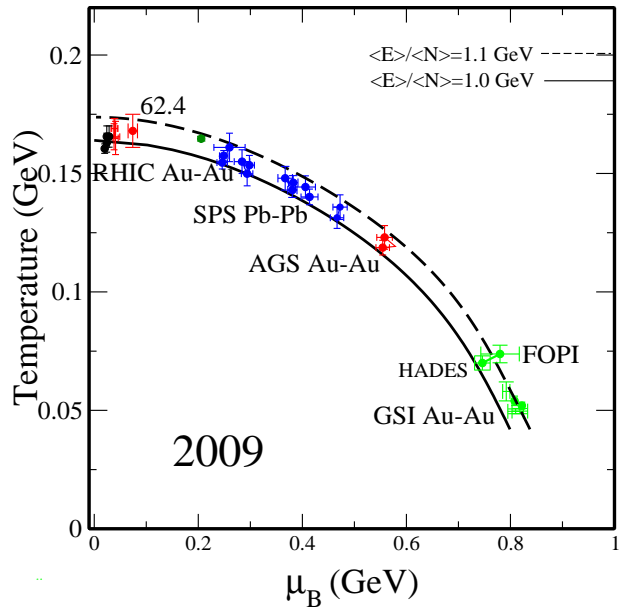}}
\caption{Chemical freeze-out temperature $T$ vs. the baryon chemical potential  at different beam 
energies together with curves corresponding to a fixed ratio of energy per hadron divided by
total number of hadrons in the resonance gas before decay of resonances.}
\label{e_2009}
\end{figure}
The resulting freeze-out curve in the $T-\mu_B$ plane can also be drawn in the
energy density vs net baryon density plane as was first done in Ref.~\cite{randrup}. The
resulting curve is shown in Fig.~\ref{randrup_figure}.
\begin{figure}[htb]
\centerline{\epsfig{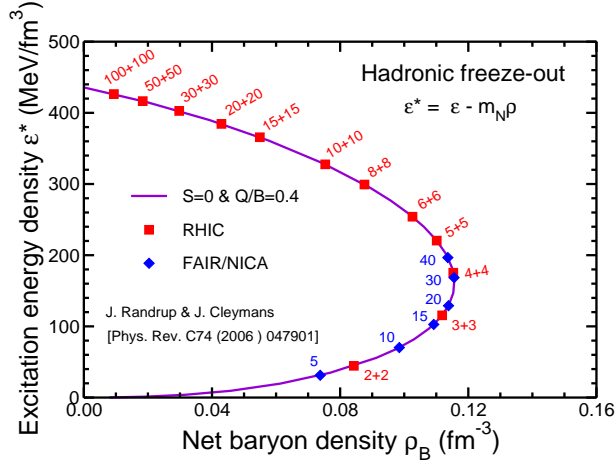}}
\caption{The hadronic freeze-out line in the $\rho_B-\eps^{*}$ phase plane 
as obtained from the values of $\mu_B$ and $T$
 that have been extracted from the experimental data in \cite{wheaton}.
The calculation employs values of $\mu_Q$ and $\mu_S$ 
that ensure $\langle S\rangle=0$ and $\langle Q\rangle=0.4\langle B\rangle$
for each value of $\mu_B$.  
Also indicated are the beam energies (in GeV/N)
for which the particular freeze-out 
conditions are expected at either RHIC or FAIR or NICA. 
}
\label{randrup_figure}
\end{figure}
This figure shows that the highest net baryon density will be reached in the beam energy covered by the NICA 
accelerator.
\section{Comparison of Chemical Freeze-Out Criteria}
\begin{figure}
\epsfig{file=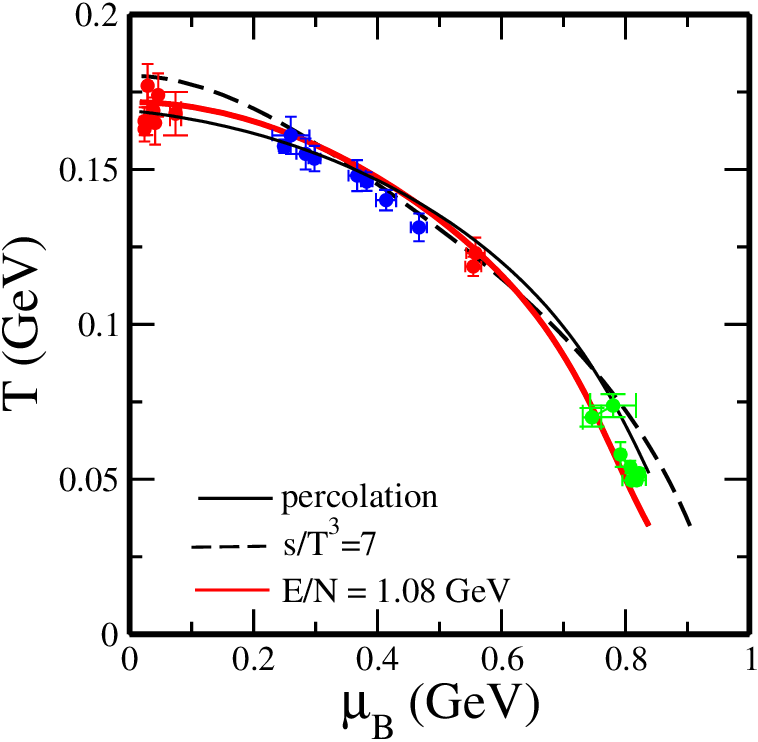,width=8cm}
\caption{Chemical freeze-out temperature $T$ vs. the baryon chemical potential  at different beam 
energies together with curves corresponding to a fixed ratio of energy per hadron divided by
total number of hadrons in the resonance gas before decay of resonances.}
\label{criteria}
\end{figure}

\begin{figure}
\epsfig{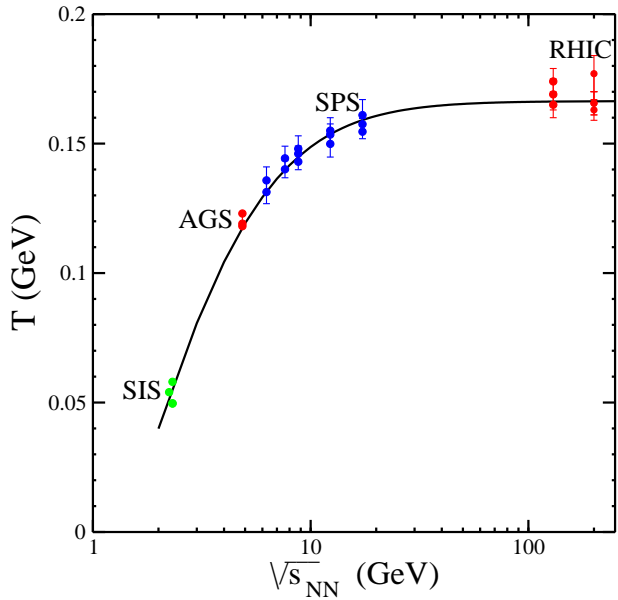}
\caption{Chemical freeze-out temperature $T$ as a function of the beam energy.}
\label{tvse}
\end{figure}

\begin{figure}
\centerline{\epsfig{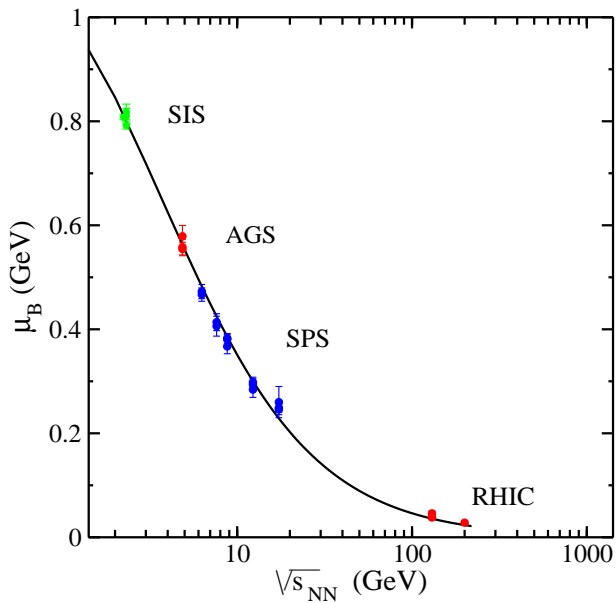}}
\caption{Chemical freeze-out baryon chemical potential $\mu_B$ as a function of the beam energy.}
\label{mubvse}
\end{figure}

\begin{figure}[htb]
\centerline{\epsfig{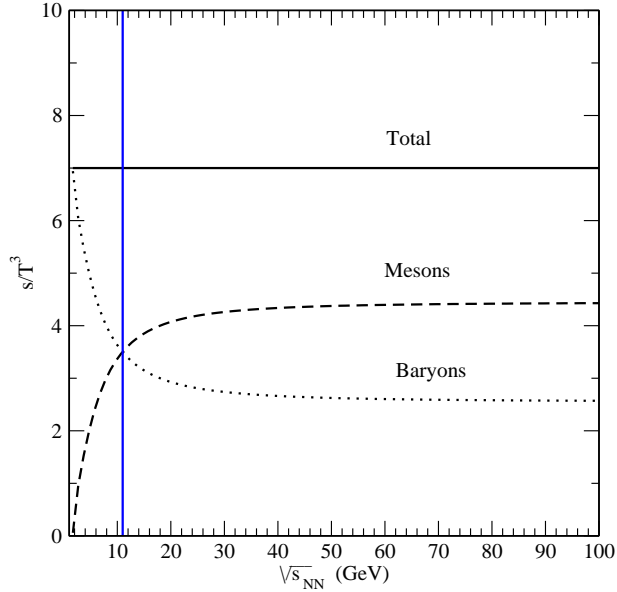}}
\caption{The $s/T^3$ ratio calculated in the thermal-statistical model along the constant value consistent with
chemical freeze-out. Also shown are the contributions from the mesons and the baryons.}
\label{sovert3}
\end{figure}

\begin{figure}[htb]
\centerline{\epsfig{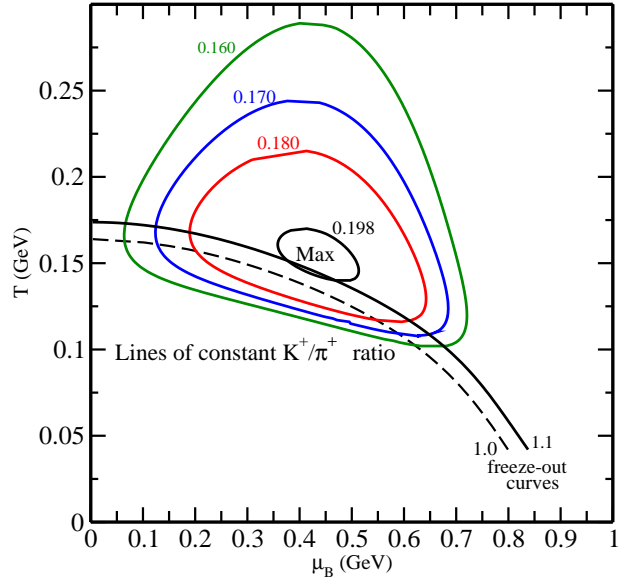}}
\caption{Lines of constant value of the $K^+/\pi^+$ ratio in the $T-\mu_B$ plane showing a clear maximum 
in this ratio close to the boundary given by the chemical freeze-out line.}
\label{kpluspiplus}
\end{figure}

In view of the success of  chemical freeze-out  in relativistic heavy ion collisions, 
much effort has gone into finding models that lead to a final state in chemical freeze-out, see e.g.  
curve~\cite{magas_satz,transition,biro}.
A comparison~\cite{wheaton} of three parameterizations is shown in Fig.~\ref{criteria}.
The corresponding dependence of the temperature and the chemical potential on beam energy is 
surprisingly smooth~\cite{wheaton} as shown in Figs.~\ref{tvse} and~\ref{mubvse}.
However, despite this smoothness in the thermal freeze-out parameters a 
roller-coaster is observed in several  particle ratios, e.g. the horn in the $K^+/\pi^+$ ratio and a similar
strong variation in the $\Lambda/\pi$ ratio~\cite{NA49}.
Again these strong variations are not observed in $p-p$ collisions and happen in the NICA energy region.
Within the framework of thermal-satistical models this variation has been connected to a change from
a baryon domicated to a meson dominated hadron gas. This conclusion is based on the observation that the entropy density
divided by the temparature to the third power, $s/T^3$, is constant over the whole energy range. The change is illustrated in Fig.~\ref{sovert3}.
Lines of constant value for the $K^+/\pi^+$ ratio are shown in Fig.~\ref{kpluspiplus} where it can be seen that the 
absolute maximum in the thermal-statistical model hugs the chemical freeze-out line. The largest observed value
is just barely compatible with this maximum. Again this is right in the energy region covered by NICA.
In the thermal-statistical model
a rapid change is expected as the hadronic gas undergoes a
transition from a baryon-dominated  to a meson-dominated gas. The
transition occurs at a temperature $T$ = 151 MeV and baryon
chemical potential $\mu_B$ = 327 MeV corresponding to an incident
energy of $\sqrt{s_{NN}}$ = 11 GeV.  
Thus the strong variation seen in the particle ratios
corresponds  to a transition from a baryon-dominated to
a meson-dominated hadronic gas. This transition occurs at a
\begin{itemize}
\item temperature $T = $ 151 MeV, 
\item baryon chemical potential $\mu_B = $ 327 MeV, 
\item  energy $\sqrt{s_{NN}} = $ 11 GeV. 
\end{itemize}
In the
statistical model this transition leads to peaks in the
$\Lambda/\left<\pi\right>$,  $K^+/\pi^+$, $\Xi^-/\pi^+$ and
$\Omega^-/\pi^+$ ratios. However, the observed ratios are sharper than the ones 
calculated in thermal-statistical models and NICA will be ideally positioned to clarify this.\\
\section{Conclusions}
There  are several theoretical  indications that the energy region 
covered by the proposed  NICA accelerator in Dubna is an extremely 
interesting one.
We present a review of data obtained in relativistic heavy ion collisions and show that there
is a gap around 11 GeV where more and better precise measurements are needed.
The theoretical interpretation can only be clarified by covering this 
energy region.
In particular the strangeness content needs to be determined, data covering
the full phase space (4$\pi$) would be very helpful to determine the thermal parameters of 
a possible phase transition and the existence of a quarkyonic phase as has been discussed in a recently~\cite{mclerran}.
\section*{Acknowledgments}
The numerous contributions by H  Oeschler,  J. Randrup, K. Redlich, E. Suhonen 
and S. Wheaton are gratefully acknowledged.

\end{document}